\documentstyle[11pt,hrd_pasp,epsfig,twoside]{article}
\begin{document}
\title{Coimbra experiment on CMD analysis: report from Group 12}
 \author{Monica Tosi}
\affil{Osservatorio Astronomico, Via Ranzani 1, 40127 Bologna, Italy}
\author{Laura Greggio}
\affil{Osservatorio Astronomico, Via Ranzani 1, 40127 Bologna, Italy;\\
 Universitaets Sternwarte Muenchen, Scheinerstrasse 1, D-81679 Muenchen, 
 Germany}
\author{Francesca Annibali}
\affil{Osservatorio Astronomico, Via Ranzani 1, 40127 Bologna, Italy}

\begin{abstract}
We describe the approach adopted by our group to derive the star formation 
history (SFH) of the chosen LMC field, as part of the experiment to compare the 
predictions obtained by different groups with the synthetic Colour-Magnitude 
diagram method. We point out what are the evolutionary characteristics of
the observed stellar populations, and present the SFH scenario which appears
to better account for them. We emphasize the importance of adopting a
{\it critical} approach when dealing with this method.

\end{abstract}

\section{Introduction}

The use of synthetic Colour-Magnitude diagrams (CMDs) allows people to 
derive the SF history of nearby galaxies by interpreting the observed 
features of their resolved stellar populations in terms of stellar 
evolution (Tosi et al 1991). Nowadays, this approach is widely followed
by the international community, but some people still wonder
whether or not the method and stellar evolution theories are reliable
enough to guarantee consistency between results obtained by 
different groups and/or procedures. The Coimbra experiment was set up to
compare the results obtained by different groups from 
the same set of high quality photometric data, treated  with the same 
procedures both for the data reduction and for the artificial star tests. 
A general report on the experiment is given by Skillman \& Gallart (this
volume, hereafter SG).

The data set refers to HST-WFPC2 images of a field in the LMC bar,
kindly made available by T. Smecker-Hane. Our group has adopted the data
catalogue provided by A. Dolphin from data reduction performed
with the HSTphot package, and his catalogue for the results of the
artificial star tests.

\section{Our approach}

To derive the SFH of the LMC field, we have applied the synthetic CMD 
method, described by Tosi et al. (this volume).
Before creating the synthetic CMDs, it is important to compare 
the empirical CMD directly with the theoretical stellar evolution 
tracks to be adopted in the synthetic CMD construction.
This comparison allows one to visualize immediately whether or
not the chosen tracks are suitable to simulate the observational CMD, and
what are the involved stellar masses, ages and metallicities. It provides
also useful information on the adoptable reddening and distance.

\begin{figure}
\plotone{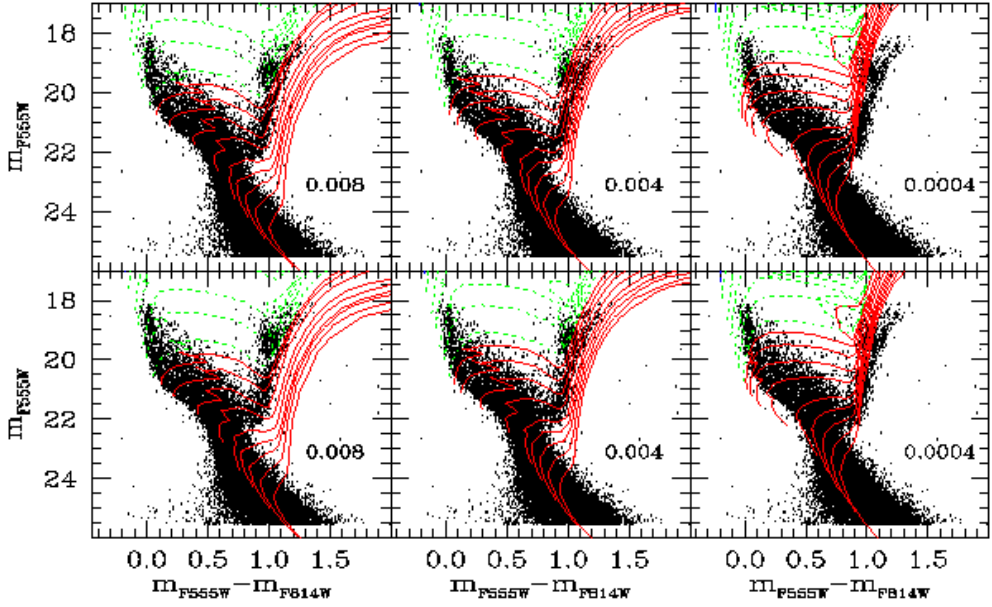}
\caption{Top panels: Stellar evolutionary tracks of the indicated metallicity
(see text for details) superimposed on the observational CMD, assuming 
$(m-M)_0=18.47$ and $E(B-V)=0.05$. Bottom panels: same as above, but assuming
$E(B-V)=0.08$. 
}
\end{figure}
\label{tracks}

The top panels of Fig.1 show the superposition of the Padova's 
stellar evolution tracks on the CMD of the LMC field, based on the reddening 
and distance modulus suggested by the experiment organizers, $(m-M)_0=18.47$ 
and $E(B-V)=0.05$, whilst the bottom panels show the effect of assuming
$E(B-V)=0.08$.
The current metallicity of the LMC medium, as derived from HII region
spectra, is Z=0.008; hence we consider this value as an upper limit to
the initial metallicity of stars born earlier than HII regions.
The shown tracks correspond to Z=0.0004 (Fagotto et al, 1994b),
Z=0.004 (Fagotto et al 1994a), and Z=0.008 (ibidem), and are those 
adopted for our experiment.  In Fig.1 models 
with M$\leq$1.9M$_{\odot}$ are plotted as solid lines, and models with 
$1.9<$M/M$_{\odot}\leq7$ are plotted as dotted lines. Stars more massive
than $\sim$4M$_{\odot}$ turn out to be absent in the CMD, both because of
a real paucity of young, bright stars and because saturated 
bright objects were removed from the photometry (see SG).

Fig.1 is instructive because it shows that with $E(B-V)$=0.05
even the most metal rich tracks are too blue to account for the observed
colour of the brighter portion of the blue plume, while $E(B-V)$=0.08 is
the minimum reddening allowing them to reach the required colours. On the
other hand, the Z=0.008 tracks have red giant branches (RGBs) much redder
than the observed one, while the Z=0.004 ones are more consistent with
the data. The Z=0.0004 models, which have the right colour at the base of the
RGB, are definitely too blue at its tip, because they show a much steeper
RGB than the other two sets. These findings let us anticipate that,
to obtain synthetic CMDs consistent with the data, we cannot have SFHs
allowing for many RGB stars with either the highest or the lowest metallicity:
most of the RGB must have metallicity around Z=0.004. 
Fig.1 indicates that also the main-sequence (MS) stars with Z=0.0004
should be a small minority of the whole old population, because the
corresponding tracks are located only at the blue edge of the faint MS
distribution, with both the adopted reddenings.

With these evolutionary constraints in mind, we have proceeded to the
construction of the synthetic CMDs. 
Following the organizers' instructions, we have assumed that the oldest
stars are 14 Gyr old; hence our simulations let the SF be already
active at that epoch.
Each extracted synthetic star was placed in the CMD by suitable 
interpolations on the Padova stellar evolution tracks of one of the
three metallicities, Z=0.0004, Z=0.004 or Z=0.008. 

The experiment data were calibrated in the ground-based Johnson-Cousins
photometric system: to transform the luminosity and temperature of the 
extracted star into {\it provisional} apparent magnitudes, we have adopted 
the bolometric corrections and colour-temperature relations from Bessel, 
Castelli, Pletz (1998) and a ratio $E(V-I)/E(B-V)$=1.25 (Dean, Warren, 
Cousins 1978).

We don't interpolate among isochrones because they are already the 
result of an interpolation which would add uncertainty on the correct
placement of the synthetic star in the CMD. Interpolation over isochrones
already converted to the observational plane would clearly add even further
numerical noise.

In each mag bin, we have retained only the fraction of extracted 
synthetic stars equal to that of recovered artificial stars in
Dolphin's artificial star catalogue. 
Then, the retained synthetic stars have been assigned a photometric error 
derived from the cumulative distribution of Dolphin's (output-input) mags 
of artificial stars with input mag equal to the {\it provisional} apparent 
mag. Since the data catalogue was cut and limited to
the range 18$\leq$V$\leq$25.5, we have applied this
same cut to the synthetic stars.
We stopped the procedure when the retained
synthetic stars were as many as the stars in the chosen portion of the
observational CMD (see Fig.2).

The synthetic CMDs were computed either with all the stars assumed to
be single, or with 30$\%$ of them assumed to be part of unresolved
binary systems with random mass ratio between the primary and 
secondary component. 

To evaluate the goodness of the model predictions, we compare them with:
the observational luminosity functions (LFs), the overall morphology of the
CMD, its mag and colour distributions, the number of objects in particular
phases (e.g. on the RGB, on the clump, on the blue loops, 
etc.). 

\section{Our results}

\begin{figure}
\plotone{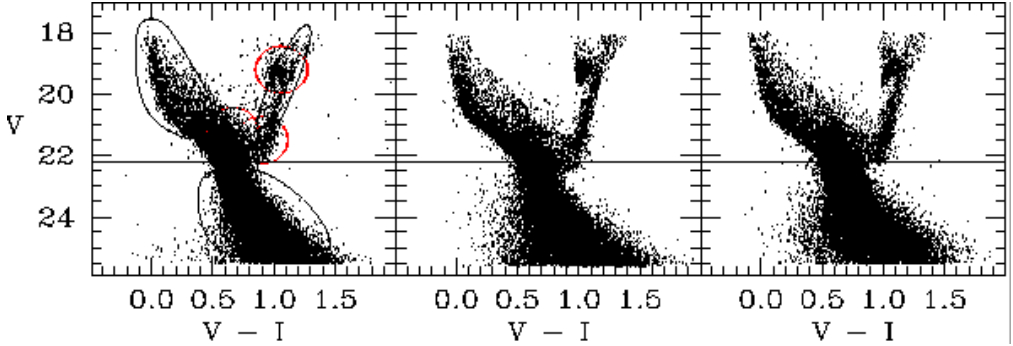}
\caption{Left-hand panel: observational CMD with superimposed the 6 
evolutionary zones to be reproduced by the models. The horizontal line
indicates the faint mag limit of the observed SGB. Central panel: synthetic
CMD with two episodes of SF, one from 14 to 3 Gyrs ago and 
the other from 3.5 Gyrs ago to now. Right-hand panel: synthetic
CMD in better agreement with the data (see text).
}
\end{figure}
\label{synth}

We have computed a variety of synthetic diagrams, letting: a) the SF range
from a constant rate through the whole galaxy lifetime to a number of
short episodes, either adjacent, or separated by quiescent phases, or
overlapping, b) the IMF exponent vary within $1.5\leq\alpha\leq2.6$
(in the units where Salpeter's $\alpha=2.35$), 
c) the metallicity in any episode be any of the three values of the adopted
tracks. For the fraction of binary stars, we have only tested the 0 and
30\% cases, because the complex morphology of the empirical CMD doesn't
allow us to really discriminate between different fractions.

To select among the computed cases, we have individuated the 6 regions of the
CMD which correspond to
evolutionary phases with well defined characteristics that can be taken as
reliable constraints. They are: the bright blue plume, the red clump, the 
RGB, the sub-giant branch, SGB, the faint MS and 
the bump at $V-I\simeq0.7$, $V\simeq21.5$;
and are indicated in the left-hand panel of Fig.2.
The fitting models are required to reproduce the number of
observed stars in each region, the luminosity functions in three colour
bins and the overall morphology of the star distribution in the CMD.

In the past (see Greggio et al 1998) we have also made use of 
bi-parametric K-S and $\chi^2$ analyses to compare the models with the data, 
but found that statistical tests are not adequate to take properly into 
account all the uncertainties related to stellar evolution, photometric 
conversions, and data calibration.
For instance, theoretical colour-temperature conversions are well known
to be particularly uncertain for the cooler stars. Synthetic CMDs are then
likely to systematically mispredict the colours of red giants.
Blind statistical tests, giving the same weight to all evolutionary phases, 
would either reject all models (including those with correct evolutionary
scenario), or retain most of them, when the tolerance is increased.
In other words, they won't be able
to discriminate between possible and impossible scenarios.
Besides, the ranking resulting from statistical tests is basically driven
by the photometric accuracy, i.e. more weight is given to the brightest
CMD regions. However, other, mostly systematic, effects impact on the
quality of the fit; for example, fluctuations due to small number statistics, 
theoretical uncertainties on the stellar models and photometric conversions, 
systematic errors in the data calibrations. In the brighter portions of
the CMD, these effects usually dominate over the random photometric errors,
so that typical statistical tests largely underestimate the real uncertainty.
As a result the model ranking is based on the fainter, more dispersed
CMD regions, where the photometric errors dominate over the systematic ones.

Given our current inability to quantify all, mostly theoretical,
uncertainties which affect the comparison, we believe that the
Hess diagram is a good tool to evidentiate the differences between 
synthetic and empirical CMDs, but also 
emphasizes the shortcomings of blind statistical tests. 

Indeed, most of the groups partecipating to the experiment who based their
results essentially on statistical tests didn't realize the need for a 
reinforcement of the SF activity in recent epochs or the need for a strong
limitation to the activity in the oldest epochs, while these results appeared
necessary to all people considering the impact on the CMD 
characteristics of specific evolutionary phases (the upper MS and the
faintest SGB, respectively).

By checking the predicted features of the 6 regions,
 we have inferred the major evolutionary characteristics of the observed
field. To summarize our results:

\noindent
{\bf 1)} Whatever the assumed 
metallicity, the SF must have been as low as possible 
at epochs earlier than about 10 Gyr ago, otherwise the field would have shown 
a significantly fainter SGB (a 14 Gyr old SGB is 0.2 mags fainter than 
the observed one, see the central panel of Fig.2). 

\noindent {\bf 2)}
A young burst, from 0.1--0.3 Gyr ago and still active with constant SF rate, is
needed to guarantee a sufficient population of the brighter portions of the 
blue plume. Adopting a flatter IMF doesn't provide the same advantage, because
the mass of the stars at the top of the MS (3--4 M$_{\odot}$) is not high 
enough to allow for a significant effectiveness of changes in the IMF slope.

\noindent {\bf 3)}
Salpeter's IMF ($\alpha=$2.35 extrapolated over the whole 0.6--120 M$_{\odot}$
range) looks appropriate for this field. Small variations in the IMF slope 
cannot be excluded, but don't seem to be needed either. 

\noindent {\bf 4)}
There is no significant signature of the presence of binary stars in the
observed CMD, but assuming in the synthetic diagrams that about 30\% of the
stars are unresolved binaries allows to more appropriately cover the whole 
color range of the   various phases: without binaries, both the bright and the 
faint MSs look too much skewed towards their blue edge.

\noindent {\bf 5)}
The lowest metallicity tracks, with Z=0.0004, are too blue to allow for
  a good fit to the data. To let them provide MS and RGB in the 
  observed colour ranges, we should assume reddenings as high as $E(B-V)=0.2$
  (and distance moduli as low as 18.17 !)

With the parameters provided by the experiment organizers, namely age
14 Gyr, $(m-M)_0=18.47$ and $E(B-V)=0.05$, we don't find a good fit to 
the data, because of the offsets already discussed above.
If we assume  $E(B-V)=0.08$ and allow for the formation of
only a few stars in the first epochs, the situation greatly improves. Our
favoured solution (shown in the right-hand panel of Fig.2)
corresponds to assuming a first episode of SF from 14 to 10 Gyr 
ago (with metallicity Z=0.004, and very low SF rate to severely limit its 
number of formed stars and make their too faint SGB not too visible), a 
second one from 10 to 3 Gyr ago, a third from 3.5 Gyr ago to the present 
epoch, and a fourth still active burst started 0.1 Gyr ago. The overlap of 
the two episodes from 3.5 to 3 Gyr ago is needed to reproduce the bump at 
$V-I=0.7$ and $V=21.5$. 

The quantitative fit achieved here depends on the adopted tracks and it is
therefore necessary to perform the experiment with other sets of models
in order to quantify the effects on the derived SFH 
of the existing uncertainties on stellar evolution theories
(e.g. Geneva vs Padova, see e.g. Greggio et al. 1998,
Aloisi et al 1999). We plan to perform this check in the
second phase of the experiment (see SG).

\section {The star formation history}

The SFH corresponding to the best synthetic CMDs described in the previous
section is shown in the bottom-right panel of Fig.3 in Tosi et al. (this 
volume). The plotted values are normalized to the 
area of the observed region and compared with the SF derived by Pagel \&
Tautvaisiene (1998) and others from studies of star cluster properties. 
The two distributions differ significantly: we want less (or no) SF at 
the earliest epochs, when they have the major peak, and we have a 
higher SF rate in the inter-peaks intervals. 
We must keep in mind, however, that their models were based on data relative
to star clusters, whereas we are studying the SF of field stars. One could
perhaps speculate that the LMC first generated the clusters and only
when their main formation was over it allowed field stars to form, at least
in our examined region. Then, around 3-4 Gyrs ago, something (interactions 
with the Milky Way ?) has triggered an enhancement in both field and cluster 
star formation. 

\acknowledgements
This work has been supported by the
Italian ASI (grant ARS-99-44) and MURST (Cofin2000).


\begin{references}
\reference Aloisi, A., Tosi, M., Greggio, L. 1999, AJ, 118, 302
\reference Bessel, M., Castelli, F., Pletz, B. 1998, A\&A 337, 321
\reference Dean, J.F., Warren, P.R., Cousins, A.W.J. 1978, MNRAS 183, 569
\reference Fagotto, F., Bressan, A., Bertelli, G., Chiosi, C. 1994, A\&AS, 
 104, 365
\reference Fagotto, F., Bressan, A., Bertelli, G., Chiosi, C. 1994, A\&AS, 
 105, 29
\reference Greggio, L., Tosi, M., Clampin, M., De Marchi, G., Leitherer, C., 
 Nota, A., Sirianni, M. 1998, ApJ, 504, 725
\reference Pagel, B.E.J. \& Tautvaisiene, G. 1998, MNRAS, 299, 535
\reference Tosi, M., Greggio, L., Marconi, G., Focardi, P. 1991, AJ, 102, 951

\end{references}
\end{document}